\documentclass[review]{elsarticle}

\usepackage{lineno,hyperref}
\usepackage{amsmath,amsfonts,amssymb}
\usepackage[ruled,linesnumbered]{algorithm2e}
\usepackage{color}
\modulolinenumbers[5]
\usepackage{multirow}
\journal{Medical Image Analysis Templates}

\usepackage{numcompress}\biboptions{authoryear}

\begin{document}

\begin{frontmatter}

\title{Momentum Contrastive Learning for Few-Shot COVID-19 Diagnosis from Chest CT Images}

\author{Xiaocong Chen\corref{cor1}\fnref{label1}}
\author{Lina Yao\fnref{label1}}
\author{Tao Zhou\fnref{label2}}
\author{Jinming Dong\fnref{label1}}
\author{Yu Zhang\corref{cor1}\fnref{label3}}
\address[label1]{School of Computer Science and Engineering at University of New South Wales, NSW 2052, Australia}
\address[label2]{Inception Institute of Artificial Intelligence, Abu Dhabi, UAE.}
\address[label3]{Department of Psychiatry and Behavioral Sciences, Stanford University, Stanford, CA 94305, USA.}
\cortext[cor1]{Corresponding authors: X. Chen. Email: xiaocong.chen@unsw.edu.au, Y. Zhang. E-mail: yzhangsu@stanford.edu}

\begin{abstract}
The current pandemic, caused by the outbreak of a novel coronavirus (COVID-19) in December 2019, has led to a global emergency that has significantly impacted economies, healthcare systems and personal wellbeing all around the world.  Controlling the rapidly evolving disease requires highly sensitive and specific diagnostics. While real-time RT-PCR is the most commonly used, these can take up to 8 hours,  and require significant effort from healthcare professionals. As such, there is a critical need for a quick and automatic diagnostic system. Diagnosis from chest CT images is a promising direction. However, current studies are limited by the lack of sufficient training samples, as acquiring annotated CT images is time-consuming. To this end, we propose a new deep learning algorithm for the automated diagnosis of COVID-19, which only requires a few samples for training. Specifically, we use contrastive learning to train an encoder which can capture expressive feature representations on large and publicly available lung datasets and adopt the prototypical network for classification. We validate the efficacy of the proposed model in comparison with other competing methods on two publicly available and annotated COVID-19 CT datasets. Our results demonstrate the superior performance of our model for the accurate diagnosis of COVID-19 based on chest CT images.
\end{abstract}

\end{frontmatter}


\section{Introduction}
The latest coronavirus, COVID-19, was initially reported in Wuhan, China toward the end of 2019 and has since spread rapidly around the globe, leading to a worldwide crisis [1]. As an infectious lung disease, COVID-19 leads to severe acute respiratory distress syndrome (ARDS) and is accompanied by a series of side effects that include 
a dry cough, fever, tiredness, shortness of breath, etc. As of May 24th 2020, more than 5.4 million individuals have been confirmed as having COVID-19, with a roughly 6.3\% case fatality rate around the world, according to the World Health Organization 
\footnote{https://www.who.int/emergencies/diseases/novel-coronavirus-2019/situation-reports} around the world.

So far, no effective treatment for COVID-19 has been found. One of the major hurdles is the lack of efficient diagnosis methods. Therefore, an accurate and rapid diagnosis platform is urgently required to conduct COVID-19 screening and prevent its further spread. Currently, most tests are based on real-time reverse transcriptase polymerase chain reaction (RT-PCR). However, each PT-PCR test can take several hours to produce results. With the current spread rate of COVID-19, this is not acceptable. Further, the limited number of test kits makes this situation even more serious ~\citep{shan+2020lung,narin2020automatic}. Recent studies show that the RT-PCR suffering from low sensitivity and low accuracy, repeated entries are required~\citep{long2020diagnosis,ai2020correlation}. This infers that patients will not be able to be confirmed on time which increases the potential risk of spreading. 

In order to address these challenges, scientists around the world are trying to develop new diagnostic systems. Some studies~\citep{bernheim2020chest,li2020coronavirus} have demonstrated that chest computed tomography (CT) imaging can help in diagnosing COVID-19 rapidly. Salehi et al.~\citep{salehi2020coronavirus} concluded that chest CT imaging is sensitive when diagnosing COVID-19 even when patients do not have clinical symptoms. Specifically, three typical radiographic features including consolidation, pleural effusion and ground class opacification can be easily observed~\citep{huang2020clinical,wang2020review,shi2020review} on COVID-19 patient’s CT images.


With this in mind, several methods based on chest CT images been developed for diagnosing COVID-19. For instance, Butt et al.~\citep{butt2020deep} adopted a convolutional neural Network (CNN) to classifier patient's CT images. In addition, there are a few works use the 3D CNN to conduct the diagnosis of COVID-19 as well based on chest CT scans~\citep{gozes2020rapid,li2020artificial,zheng2020deep}. Mei et al.~\citep{mei2020artificial} adopted the ResNet to conduct the rapid diagnosis of the COVID-19. Besides the diagnosis, lots of works are using the segmentation technique to conduct the detection~\citep{chen2020deep,shi2020large,chen2020residual}. All those existing methods are trained based on the limited available samples which have small number of patients and may not have capability to generalize to new patients.
It is well-known that, the lack of labelled training data is the common problem, while deep learning based methods generally require a large volume of data to accurately train the models. Many research efforts have been sought for alleviating this problem such as data augmentation or through generative adversarial network (GAN)~\citep{zhao2019data,akkus2017deep,oliveira2017augmenting,pereira2016brain}.However, these methods are highly sensitive with the parameter selection. Hand-tuned data augmentation methods like rotation may lead to over-fitting~\citep{eaton2018improving} and the generated images by GAN can not simulate the real patient which may introduce unpredictable bias in testing phase~\citep{zhao2019data}. Recently, few-shot learning has been attracting much attention in medical image analysis. In general, few-shot learning aims to leverage existing data to classify new tasks from similar domains. The basic workflow for few-shot learning is first pre-training an embedding network on a large dataset (e.g. ImageNet) and then fine-tuning the weights of this network, finally applying it into a unseen small dataset ~\citep{he2020sample}. 
However, the performance is limitedly improved. One reason lies in the ImageNet contains a broad range of categories of images and pre-train on ImageNet usually might bring in unrelevant information, so as not to learn a effective embeddings for improving lung-specific feature representation. On the other hand, pre-training on ImageNet causes high computational cost, for example, ImageNet-1B normally required more than 50 GPU days.


To address this challenge, we develop an end-to-end trainable deep few-shot learning framework to make an accurate prediction with minimal training Chest CT Images. Specifically, we fist use the instance discrimination task to enforce model to discriminate two images are the same instance or not. We generate different views of the same images to augment the original dataset. As the goal at this stage is increasing variances other than discrimination, we can avoid the disadvantages of data augmentation mentioned previously.



We then deploy a self-supervised strategy~\citep{oord2018representation} powered with momentum contrastive training to further boost the performance.  
The key idea is to build a dynamic dictionary to perform key, query look-up where the keys are sampled from data and encoded by the encoder. However, the key in dictionary is noisy and inconsistent due to the back-propagation~\citep{he2019momentum}. The momentum mechanism is applied to mitigate this effect by updating key encoder and query encoder in different scales. Finally, we utilized two public lung datasets to pre-train an embedding network and employ the prototypical network~\citep{snell2017prototypical} to conduct the few-shot classification, which learns a metric space where the classification can be performed by measuring the distances to the derived prototypical representation of each class. 
The extensive experiments on two new datasets demonstrate that our model provides a promising tool for quick COVID-19 diagnosis with very limited available training data. 

\section{Problem Definition}
Due to the shortage of the annotated COVID-19 CT images, the normal classification methods may not able to work properly. Based on that, we formulate the COVID-19 diagnosis problem as a few-shot classification problem. The few-shot learning is designed for the case which only a few samples available in a new class on a classification task.
The few-shot learning can be defined as a $M$-way, $C$-shot episodic task~\citep{vinyals2016matching} where $M$ represents the number of classes and $C$ represents number of samples available for each class. The training set which never seen before can be represented as $D=\{(x_0,y_0),\cdots,(x_d,y_M)\}$, $d$ is the number of samples in this dataset. We randomly selected the support set and query set from $D$: i) The support set $S$ can be partially or fully made up of $M$ classes but only contain $C+1$ samples each. ii) We randomly select 1 sample from the $C+1$ samples to form the test set (query set). Hence, COVID-19 diagnosis can be represented as a two-way, $C$-shot learning problem.

\section{Methodology}
\begin{figure}[!h]
    \includegraphics[width=\linewidth]{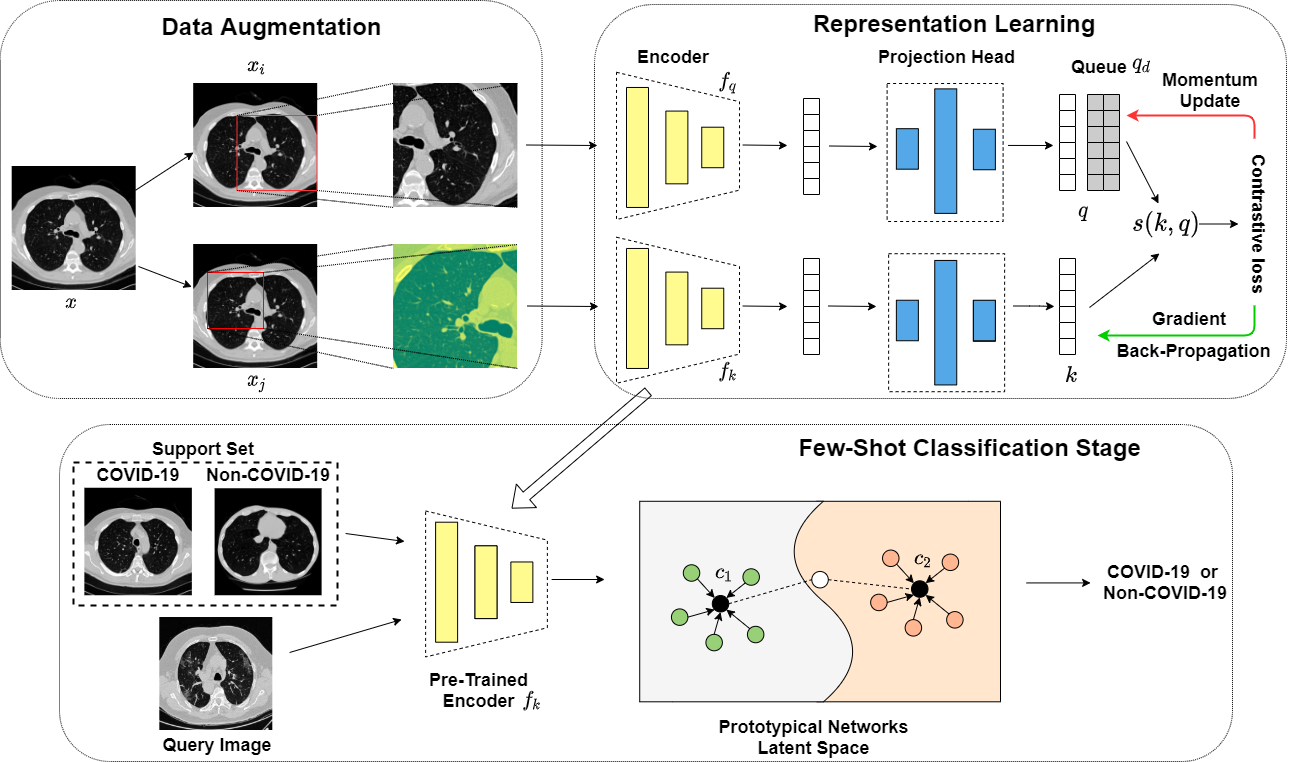}
    \caption{The overall architecture of our approach. Top: The pre-training stage, which includes data augmentation and representation learning. The pretext task is an instance discrimination task. Bottom: Few-shot classification with 2-way, 1-shot example. For classification, the support images and query image are encoded by the pre-trained encoding network. Query sample embeddings are compared with the centroid of training sample embeddings and fine-tuning the pre-trained encoder.}
    \label{fig:structure}
\end{figure}

In this section, we will introduce our proposed self-supervised COVID-19 diagnosis method. The overall flowchart is illustrated in Fig. \ref{fig:structure}. We will describe the three major components of our method which include data augmentation, representation learning and the few-shot classification.
\subsection{Data Augmentation}
\begin{figure}[h]
    \centering
    \includegraphics[width=\linewidth]{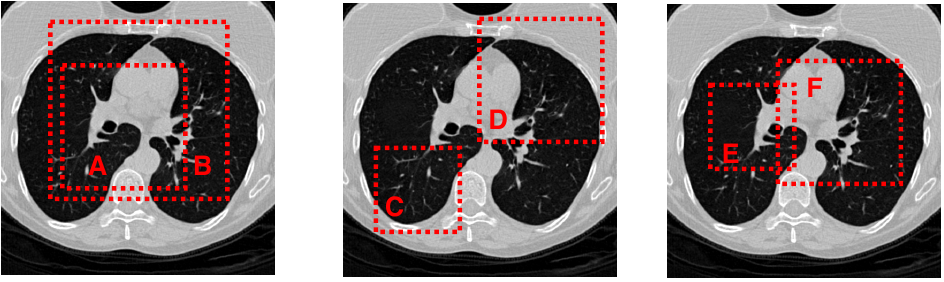}
    \caption{Three possibilities for random cropping. Dashed boxes are augmented views. Crops A, C, E views will have random color distortions applied while B, D, F will not change if the method (2) is chosen. All the cropped sections will be resized back to the original input image size. For the instance discrimination task, the goal, given B, D, F,is to, determine whether or not A, C, E are in the same instances.}
    \label{fig:aug}
\end{figure}

Data augmentation has been widely used in unsupervised representation learning and supervised learning~\citep{parkhi2012cats,donahue2019large,donahue2014decaf}. A few existing approaches define the contrastive classification task as changing abd image's structure. For instance, Hjelm et al.~\citep{hjelm2018learning} and Bachman et al.~\citep{bachman2019learning} used global-to-local view for contrastive learning as shown in the first example in Fig~\ref{fig:aug}. Meanwhile, Oord et al.~\citep{oord2018representation} and Henaff et al.~\citep{henaff2019data} achieved neighbor prediction using the adjacency view (middle example, Fig~\ref{fig:aug}). Those two methods can be view on the first two images on \ref{fig:aug}. Besides, we also introduce the over-lapping view which is the third image on Fig \ref{fig:aug}.

In this study, we apply a stochastic data augmentation $\mathcal{T}$ which will randomly transfer a given example image $x$ into two different views denoted as $x_i,x_j$. We consider the pair $x_i,x_j$ as positive. In this study, we apply two simple augmentation strategies in sequence: 1) random cropping, followed by a resizing operation back to the original size with random flipping. 2) random cropping with color distortions followed by a resizing operation. When a new image is fed into the model, one of the above methods to perform is randomly selected for augmentation. This process is repeated twice to generate two different views.

Note that we implement color distortions using the  torchvision\footnote{https://pytorch.org/docs/stable/torchvision/index.html} package in PyTorch~\citep{paszke2019pytorch}.
\subsection{Contrastive Visual Embedding}
Using contrastive learning to learn visual embeddings was first explored by Hadsell et al.~\citep{hadsell2006dimensionality}. The task can be defined as: Given an image set $\{\mathcal{I} = i_1,\cdots,i_p\},x_i\in\mathbb{R}^d$, the goal is to find a mapping function $G: \mathbb{R}^d \mapsto \mathbb{R}^a, a\ll d$ to satisfy:
\begin{align}
    s(G(x),G(x^+) \gg s(G(x),G(x^-)
\end{align}
where $s(\cdot,\cdot)$ is a function used to measure the similarity between two inputs. $G$ is designed for dimension reduction and representation learning. Finally, $x^+,x^-$ represent the positive and negative samples which means $x^+$ being similar to $x$ and $x^-$ dissimilar. It is worth to mentioned that, the contrastive learning is a type of unsupervised learning.
A simple framework for the contrastive learning was proposed by Chen et al.~\citep{chen2020simple}. The framework learns the representations by maximizing the agreement between differently augmented views $x_i, x_j$ of the same data example $x$ via a contrastive loss in the latent space. We adopt this framework in our model. Specifically,our representation learning stage consists of three modules: encoder, projection head and the contrastive loss function.

\noindent{\textbf{Encoder.}}
The neural network based encoder $f(\cdot)$ can extract representations from the augmented images. This framework is flexible to adopt any type of network architecture without constraints. In this study, we adopt ResNet~\citep{he2016deep} to obtain the representation $h_i, h_j$, $h_i = f(x_i) = \text{ResNet}(x_i)$ where $h_i$ is the $\mathbb{R}^d$ output of the average pooling layer. \\
\textbf{Projection Head.}
The project head $g(\cdot)$ is a function that can map the resulting representation int oapplication space of the contrastive loss. The most common projection head used is the multilayer perceptron (MLP) with one hidden layer~\citep{chen2020simple,chen2020improved}. In this case, we can express the $z_i$ (as well as $z_j$) as:
\begin{align}
    z_i = g(h_i) = W^2\sigma(W^1 h_i)
\end{align}
where $W^1, W^2$ are the weights of the hidden layer and output layer respectively. The $\sigma(\cdot)$ is the non-liner ReLU activation function which can be defined as:
\begin{align}
    \text{ReLU}(x) = 
    \begin{cases} 
      0 & x\leq 0 \\
      x & x > 0
   \end{cases}
\end{align}
We will examine the effectiveness of this projection head in Section \ref{sub:exp}.\\
\textbf{Contrastive Loss Function.}
The contrastive loss function is defined for the contrastive pre-text task. We only consider the instance discrimination task~\citep{wu2018unsupervised} in this study. Given a set $\{x_k\}$ including a positive pair $x_i,x_j$, the contrastive task aims to identify $x_j$ in set $\{x_k\}_{k\neq i}$ for a given $x_i$.\\
We define the contrastive task on pairs of augmented images from a randomly selected minibatch with $N$ samples. The augmentation process results in $2N$ data points. To create the contrastive task, we need enough negative samples to construct the loss function. Similar to Chen et al.~\citep{doersch2017multi}, we treat the other $2N - 2$ examples as the negative samples. The similarity function $s(\cdot,\cdot)$ can be defined as the cosine similarity:
\begin{align}
    s(v,u) = \frac{v^\intercal u}{\|v\|\|u\|}
\end{align}
where $v,u$ are two vectors. Based on this, we can define the loss function for a pair of positive samples $(i,j)$ as:
\begin{align}
    \mathcal{L}_{i,j} = -\log\frac{\exp(s(z_i,z_j)/\tau)}{\sum_{k=1}^{2N}\mathfrak{1}_{k\neq i}\exp(s(z_i,z_k)/\tau)}
\end{align}
Here $\mathfrak{1}_{k\neq i}\in\{0,1\}$ is the indicator which have value of 1 when $k\neq i$ and 0 otherwise, and $\tau$ is the temperature parameter. This loss is known as the normalized temperature-scaled cross entropy loss~\citep{sohn2016improved,bachman2019learning,oord2018representation}.
However, Eq.(5) only considers the positive samples and ignores negative samples. This may lead to potential bias. To avoid this, we introduce the momentum mechanism into our model. 

Contrastive learning can also be expressed as training an encoder to conduct a dictionary lookup task. Consider an encoded query $q$ and encoded samples $x_i,...,x_k$ which are the keys of the dictionary. If the query $q$ is similar to the sample $x^+$ this means there is a match. For the negative samples $x^-$, there is no match in the dictionary. Based on this definition, He et al. proposed an unsupervised learning-based framework MoCo~\citep{he2019momentum}, by adopting contrastive learning.

Based on the above definition, the goal of contrastive learning is to build a discrete dictionary for high-dimensional continuous inputs. The core of  MoCo is maintaining a dictionary with a queue. The benefit of this is that the encoder can reuse the encoded keys from the previous mini-batch. In addition, the dictionary can be much larger than the mini-batch and easy to adjust. As the number of samples that can be included in the dictionary is fixed, once the dictionary is full, it will progressively remove the oldest records. In this way, the consistency of the dictionary can be maintained as the oldest samples are often out-of-date and inconsistent with the new entries. Another approach, called Memory Bank~\citep{wu2018unsupervised}, tries to store the historical records of the encoded samples. This approach maintains a bank of all the representations of the dataset. The dictionary then randomly samples from the memory bank directly for each mini-batch without back-propagation. However, this method will lead to inconsistency when sampling. To overcome this, back-propagation should be conducted to keep the sampling step up-to-date. A simple solution is to copy the key encoder $f_k$ from query encoder $f_q$ without the gradient. However, the encoder changes constantly which can lead to a noisy key representation and poor results. The momentum contrast was address to relief this problem, using a different method to update the gradient for $f_k$:
\begin{align}
    \theta_k \xleftarrow{} m\theta_k + (1-m)\theta_q.
\end{align}
where $\theta_k$ is the parameter for $f_k$, $\theta_q$ is the parameter for $f_q$ and $m\in[0,1)$ is the momentum coefficient. We use the back-propagation to update the parameter $\theta_q$ and use Eq.(6) to update $\theta_k$. Benefiting from the momentum coefficient, the $\theta_k$'s update is smoother than $\theta_q$. Based on the different update strategies, the query and key will be encoded by different encoders eventually.

Based on the above discussion, we use the dictionary as a queue to allow the encoder to reuse the previous encoded sample. The loss function for the pre-trained model can be written as:
\begin{align}
    \mathcal{L} = -\log\frac{\exp(q,k^+)/\tau}{\exp(q,k^+)/\tau + \sum_{k^-}\exp(q,k^-)/\tau}.
\end{align}
Different from Eq.(5), here we need to consider the queue and the negative cases, so we slightly modify the loss function to fulfil this requirement by introducing the positive examples $k^+$ and negative examples $k^-$, where $q_k = k^+ \cup k^-$. In the instance discrimination pre-text task, a positive pair is formed when a query $q$ and a key $k$ are augmented from the same sample; otherwise, a negative pair is created. Once the pre-training step is finished, we extract the pre-trained encoder $f(\cdot)$ and integrate it into our classification module.

\subsection{Prototypical Network for Few-Shot Classification}
Another step in our workflow is classification. In this stage, meta-learning is applied to fine-tune the pre-trained encoder to fit the class changes required by few-shot learning. 
Then we use the Prototypical Networks~\citep{snell2017prototypical} to conduct the few-shot classification. The prototypical network learns an embedding that maps all inputs into a mean vector $c$ in the latent space to represent
each class. The goal of the pre-trained encoder is to similar images close and dissimilar images separate in the latent space. The prototypical network has a similar aim so it to fine-tune our pre-trained encoder. For class $m$, the centroid embedding features can be written as:
\begin{align}
    c_m = \frac{1}{|S|}\sum_{(x_d,y_M)\in S} \psi(x_d)
\end{align}
where $\psi(\cdot)$ is the embedding function from the prototypical network. As the prototypical network is a metric based learning method, we use the Euclidean distance to produce the distribution for all classes for a query $q$.
\begin{align}
    p(y=m|q) = \frac{\exp(-d(\psi(q),c_m))}{\sum_{m'}\exp(-d(\psi(q),c_{m'})}.
\end{align}
Eq.(9) is based on the softmax function over the distance between a query set's embedding and the features of the class. The loss function for this stage can be defined as:
\begin{align}
    \mathcal{L_{\text{meta}}} = d(\psi(q),c_m) + \log (d(\psi(q),c_{m'}))
\end{align}
\subsection{Training Strategy}
Algorithm \ref{alg:train} shows the whole pre-training workflow of our model.
\begin{algorithm}[!h]
\SetKwInOut{Input}{input}\SetKwInOut{Output}{output}
\SetAlgoLined
\Input{Batch size $N,\tau,f_k,f_q,g,\mathcal{T},q_k$}

 \For{\text{sampled mini-batch} $\{x_k\}_{k=1}^N$}{
    \For{$k \in \{1,\cdots,N\}$}{
        Select two data augmentation functions from $\mathcal{T}$: $t,t'$ \;
        $x_{2k-1} = t(x_{k}), \widehat{x}_{2k-1} = t'(x_{k})$ \;
        $h_{2k-1} = f_k(x_{2k-1}), h_{2k} = f_q(\widehat{x}_{2k-1})$ \;
        $z_{2k-1} = g(h_{2k-1}), z_{2k} = g(h_{2k})$ \;
    }
    \For{$i \in \{1,\cdots,2N\}, j\in  \{1,\cdots,2N\}$}{
        Calculate the similarity using Eq.(4) \;
    }
    Update $f_k$ to minimize Eq.(7)\;
    Update $f_q$ by Eq.(6) \;
    enqueue($q_k$, $z_{2k-1}$) \;
    dequeue($q_k$)\;
 }
 \textbf{return} $f_k$\;
\caption{Training algorithm for the pre-training}
\label{alg:train}
\end{algorithm}

\section{Experiments}
\label{sub:exp}
\subsection{Datasets}

We evaluate our proposed model using two publicly available annotated COVID-19 CT image datasets: (1) COVID-19 CT provided by Zhao et al.~\citep{zhao2020COVID-CT-Dataset} and (2) a dataset provided by the Italian Society of Medical and Interventional Radiology\footnote{https://www.sirm.org/category/senza-categoria/covid-19/}, processed by MedSeg\footnote{http://medicalsegmentation.com/covid19/}. The proposed model requires proper pre-training. Different from existing methods like Self-Trans~\citep{he2020sample} that use the ImageNet to pre-train the model, we use DeepLesion~\citep{yan2018deep} and the Lung Image Database Consortium Image Collection (LIDC-IDRI)\footnote{https://wiki.cancerimagingarchive.net/display/Public/LIDC-IDRI} to pre-train our model. DeepLesion contains over 32,000 lung CT images, while LIDC-IDRI has 244,617.
Both datasets are public datasets and focus on lung diseases. We use those two datasets without labels to pre-train the encoder network.

When dividing the support and query sets for classification, we divide the dataset at a patient-level instead of CT level to avoid any possible over-fitting. We report the basic statistics for the COVID-19 CT dataset and MegSeg in Table \ref{tab:stat}.
\begin{table}[h]
    \centering
    \caption{Number of patients and number of CT images available in test datasets.}
    \begin{tabular}{|c|c|c|c|}
        \hline
                                 &              & COVID-19 CT & MegSeg \\ \hline
\multirow{2}{*}{\# of Patients}  & COVID-19     & 216         & 43     \\ \cline{2-4} 
                                 & Non-COVID-19 & 171         & 0      \\ \hline
\multirow{2}{*}{\# of CT images} & COVID-19     & 349         & 110    \\ \cline{2-4} 
                                 & Non-COVID-19 & 397         & 0      \\ \hline
     \end{tabular}
    \label{tab:stat}
\end{table}\\
We combine the two datasets for testing. Note that all CT images were resized to 512 $\times$ 512 using opencv2\footnote{https://opencv.org/}.
\subsection{Experimental Settings}
For pre-training, we use the SGD optimizer with a weight decay of 0.0001 and momentum of 0.9. The momentum update coefficient is 0.999. The mini-batch size is set to 256 in eight GPUs. The initial learning rate is 0.03. The number of epochs is 200, and the learning rate is multiplied by 0.1 after 120 and 160 epochs, as described in ~\citep{wu2018unsupervised}. The encoder is ResNet-50. The two-layer MLP projection head has a 2048-D hidden layer with a ReLU activation function. The weights are initialized using He initialization~\citep{he2015delving}, and the temperature parameter $\tau$ is set to 0.07. For the classification stage, we follow the default settings of the prototypical net. The experiments were conducted on eight GPUS which includes six NVIDIA TITAN X Pascal GPUs and two NVIDIA TITAN RTX. 
\subsection{Evaluation and Results}
We evaluate our approach by using four metrics: i) Accuracy, which measures the percentage of correctly classified samples over the whole dataset; ii) Precision, which measures the percentage of true positives (TP) over all predicted positive samples; iii) Recall, used to measure the percentage of TPs over all positive samples; and iv) Area-under-the-curve (AUC) which measures the relation between FPs and TPs. For fair comparison, we train and test all baseline methods on COVID-19 CT and MegSeg dataset directly by using 10-fold cross-validation at a patient-level with cross-entropy loss function.
\begin{figure}[!h]
    \centering
    \includegraphics[width=\linewidth]{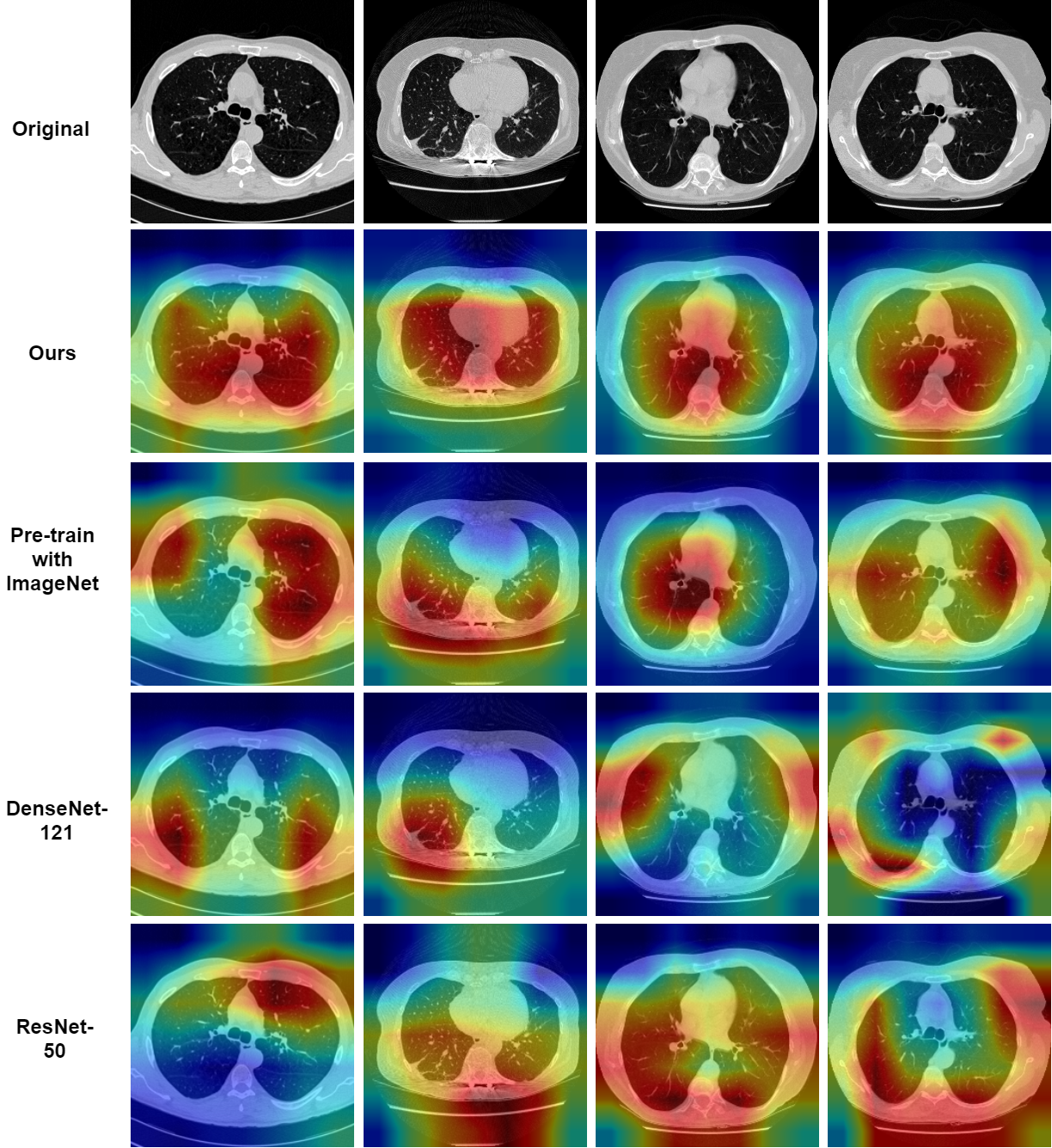}
    \caption{Grad-CAM~\citep{selvaraju2017grad} visualizations for learned features from several baseline methods. Top line is the original image set, followed by our methods, our methods with ImageNet pre-train, DenseNet-121 and ResNet-50. It is obviously that our method can learn lung's features better.}
    \label{fig:features}
\end{figure}
The results are shown in in Table~\ref{tab:res}. We find that two-way, one-shot method achieves very similar result to the ResNet-50 trained on the COVID-19 dataset. In addition, we also provide the visualization of the learned feature on figure\ref{fig:features}. 
\begin{table}[h]
    \centering
    \caption{Comparison Results between the proposed model and other methods. Our method uses a two-way, one-shot strategy.}
    \begin{tabular}{c|c|c|c|c}
         \hline
         & Accuracy & Precision & Recall & AUC \\ \hline
         ResNet-50 & 0.873 $\pm$ 0.013 & 0.894 $\pm$ 0.012 & 0.874 $\pm$ 0.012 & 0.935 $\pm$ 0.014\\ \hline
         ResNet-12 & 0.834 $\pm$ 0.014 &0.851 $\pm$ 0.013 & 0.844 $\pm$ 0.014 & 0.901 $\pm$ 0.015 \\ \hline
         DenseNet-121 & 0.855 $\pm$ 0.013 & 0.867 $\pm$ 0.012 & 0.859 $\pm$ 0.012 & 0.894 $\pm$ 0.013 \\ \hline
         Ours & 0.868 $\pm$ 0.012 & 0.883 $\pm$ 0.011 & 0.872 $\pm$ 0.012 & 0.931 $\pm$ 0.013 \\ \hline
    \end{tabular}
    \label{tab:res}
\end{table}
As aforementioned, our method is based on few-shot learning. We are interested in how the number of shots affect model's performance. Based on that, we conduct a few experiments to explore the relationship between the performance and the number of shots. We use the ResNet-50 as the baseline methods. The results are shown in Table~\ref{tab:res_2}.
\begin{table}[h]
    \centering
    \caption{Results for different settings. Here, $W$ indicates the number of ways and $S$ the number of shots. For instance, 2W1S represents the two way, one shot setting.}
   \begin{tabular}{c|c|c|c|c}
         \hline
         & Accuracy & Precision & Recall & AUC \\ \hline
         ResNet-50 &  0.873 $\pm$ 0.013 & 0.894 $\pm$ 0.012 & 0.874 $\pm$ 0.012 & 0.935 $\pm$ 0.014 \\ \hline
         Ours(2W,1S) & 0.868 $\pm$ 0.012 & 0.883 $\pm$ 0.011 & 0.872 $\pm$ 0.012 & 0.931 $\pm$ 0.013 \\ \hline
         Ours(2W,2S) & 0.872 $\pm$ 0.012 & 0.890 $\pm$ 0.011 & 0.875 $\pm$ 0.012 & 0.935 $\pm$ 0.012 \\ \hline
         \textbf{Ours(2W,3S)} & \textbf{0.876 $\pm$ 0.012} & \textbf{0.895 $\pm$ 0.011 } & \textbf{0.878 $\pm$ 0.011} & \textbf{0.938 $\pm$ 0.12}  \\ \hline
         \textbf{Ours(2W,4S)} & \textbf{0.881 $\pm$ 0.012} & \textbf{0.898 $\pm$ 0.011 } & \textbf{0.882 $\pm$ 0.011} & \textbf{0.942 $\pm$ 0.012}  \\ \hline
    \end{tabular}
    \label{tab:res_2}
\end{table}\\
It is not hard to find that when the number of shots increasing, the performance increasing gradually. Specifically, when the number of shots is larger than 3, the result of our model is better then that of ResNet-50. These results show indicate that the pre-trained encoder can capture the features from unknown images very well.
\subsection{Ablation Study}
In this section, we conduct the ablation studies to demonstrate the effectiveness of each component. The default setting is two-way, one-shot and the ResNet-50 use the same setting as previous section. In summary, we want to answer the following research questions: 1) What if pre-trained was conducted on ImageNet? 2) How data augmentation and projection head affect the performance? 3)How important of the fine-tuning stage?\\
First, we are interested in the pre-training dataset selection. We thus conduct the experiments again using ImageNet for pre-training on default setting. The results can be found in Table~\ref{tab:img}.
\begin{table}[h]
    \centering
    \caption{PreAblation study results for pre-training dataset selection. Both models use the two-way, one-shot setting.}
    \begin{tabular}{c|c|c|c|c}
        \hline
         & Accuracy & Precision & Recall & AUC \\ \hline
         Ours & 0.868 $\pm$ 0.012 & 0.883 $\pm$ 0.011 & 0.872 $\pm$ 0.012 & 0.931 $\pm$ 0.013  \\ \hline
         ImageNet & 0.732 $\pm$ 0.023 & 0.744 $\pm$ 0.021 & 0.738 $\pm$ 0.023 & 0.870 $\pm$ 0.022 \\ \hline
    \end{tabular}
    \label{tab:img}
\end{table}
As expected, we find that the performance is worse when the model is pre-trained by ImageNet. As discussed before,  an extra step may be required to conduct transfer learning from common items to lung CT images.

Next, we discover the role of the data augmentation and projection head played in our model. We've conducted the experiments on our model without projection head and without augmentation. The result can be found in Table~\ref{tab:aug}.
\begin{table}[h]
    \centering
    \caption{Data augmentation and projection head dffect}
    \begin{tabular}{c|c|c|c|c}
        \hline
         & Accuracy & Precision & Recall & AUC \\ \hline
         Ours & 0.868 $\pm$ 0.012 & 0.883 $\pm$ 0.011 & 0.872 $\pm$ 0.012 & 0.931 $\pm$ 0.013  \\ \hline
         No Aug.&  0.779 $\pm$ 0.021 & 0.791 $\pm$ 0.020  & 0.780 $\pm$ 0.021 & 0.889 $\pm$ 0.022 \\ \hline
         No Proj. & 0.856 $\pm$ 0.013  & 0.875 $\pm$  0.012 & 0.870 $\pm$ 0.013  & 0.910 $\pm$ 0.015 \\ \hline
    \end{tabular}
    \label{tab:aug}
\end{table}
As we can see, the data augmentation have the significant effect with the result and the projection head can boost the performance.
Finally, we want to examine the effect of the fine-tuning process. To investigate this problem, we first pre-train the embedding network and make some modifications with the few-shot classification stage by replace the prototypical network with frozen features liner classifier. The frozen features linear classifier means we directly apply the linear classifier into the learned embedding network without any weights update procedure. Result can be found in Table~\ref{tab:ft}.
\begin{table}[h]
    \centering
    \caption{Effect of fine-tuning}
    \begin{tabular}{c|c|c|c|c}
        \hline
         & Accuracy & Precision & Recall & AUC \\ \hline
         Ours & 0.868 $\pm$ 0.012 & 0.883 $\pm$ 0.011 & 0.872 $\pm$ 0.012 & 0.931 $\pm$ 0.013  \\ \hline
         Linear Classifier & 0.788 $\pm$ 0.011  & 0.795 $\pm$ 0.010 & 0.790  $\pm$ 0.010 & 0.892  $\pm$ 0.009 \\ \hline
    \end{tabular}
    \label{tab:ft}
\end{table}

\section{Conclusion}
Nowadays, CT imaging attracts more and more attention as a screening tool for diagnosing COVID-19. It provides a visualization for community to monitor patient's progression and can help to evaluate the severity of COVID-19~\citep{shan+2020lung}. However, the lack of the annotated CT scans are the biggest challenge. In this study, we proposed a new deep-learning based method which can be used for automatic screening of COVID-19 with limited samples. And it proved that such method achieves superior performance than ResNet-50 when the number of available samples is larger than three. ResNet is a well-known and widely used supervised learning model on medical image area. As a self-supervised method which belongs to unsupervised learning field, the results are better than ResNet would be remarkable.

\bibliography{mybibfile}

\begin{thebibliography}{47}
\expandafter\ifx\csname natexlab\endcsname\relax\def\natexlab#1{#1}\fi
\providecommand{\url}[1]{\texttt{#1}}
\providecommand{\href}[2]{#2}
\providecommand{\path}[1]{#1}
\providecommand{\DOIprefix}{doi:}
\providecommand{\ArXivprefix}{arXiv:}
\providecommand{\URLprefix}{URL: }
\providecommand{\Pubmedprefix}{pmid:}
\providecommand{\doi}[1]{\href{http://dx.doi.org/#1}{\path{#1}}}
\providecommand{\Pubmed}[1]{\href{pmid:#1}{\path{#1}}}
\providecommand{\bibinfo}[2]{#2}
\ifx\xfnm\undefined \def\xfnm[#1]{\unskip,\space#1}\fi
\bibitem[{Ai et~al.(2020)Ai, Yang, Hou, Zhan, Chen, Lv, Tao, Sun and
  Xia}]{ai2020correlation}
\bibinfo{author}{Ai\xfnm[ T.]}, \bibinfo{author}{Yang\xfnm[ Z.]},
  \bibinfo{author}{Hou\xfnm[ H.]}, \bibinfo{author}{Zhan\xfnm[ C.]},
  \bibinfo{author}{Chen\xfnm[ C.]}, \bibinfo{author}{Lv\xfnm[ W.]},
  \bibinfo{author}{Tao\xfnm[ Q.]}, \bibinfo{author}{Sun\xfnm[ Z.]},
  \bibinfo{author}{Xia\xfnm[ L.]}.
\newblock \bibinfo{title}{Correlation of chest ct and rt-pcr testing in
  coronavirus disease 2019 (covid-19) in china: a report of 1014 cases}.
\newblock \bibinfo{journal}{Radiology}
  \bibinfo{year}{2020};:\bibinfo{pages}{200642}.
\bibitem[{Akkus et~al.(2017)Akkus, Galimzianova, Hoogi, Rubin and
  Erickson}]{akkus2017deep}
\bibinfo{author}{Akkus\xfnm[ Z.]}, \bibinfo{author}{Galimzianova\xfnm[ A.]},
  \bibinfo{author}{Hoogi\xfnm[ A.]}, \bibinfo{author}{Rubin\xfnm[ D.L.]},
  \bibinfo{author}{Erickson\xfnm[ B.J.]}.
\newblock \bibinfo{title}{Deep learning for brain mri segmentation: state of
  the art and future directions}.
\newblock \bibinfo{journal}{Journal of digital imaging}
  \bibinfo{year}{2017};\bibinfo{volume}{30}(\bibinfo{number}{4}):\bibinfo{pages}{449--459}.
\bibitem[{Bachman et~al.(2019)Bachman, Hjelm and
  Buchwalter}]{bachman2019learning}
\bibinfo{author}{Bachman\xfnm[ P.]}, \bibinfo{author}{Hjelm\xfnm[ R.D.]},
  \bibinfo{author}{Buchwalter\xfnm[ W.]}.
\newblock \bibinfo{title}{Learning representations by maximizing mutual
  information across views}.
\newblock In: \bibinfo{booktitle}{Advances in Neural Information Processing
  Systems}. \bibinfo{year}{2019}. p. \bibinfo{pages}{15509--15519}.
\bibitem[{Bernheim et~al.(2020)Bernheim, Mei, Huang, Yang, Fayad, Zhang, Diao,
  Lin, Zhu, Li et~al.}]{bernheim2020chest}
\bibinfo{author}{Bernheim\xfnm[ A.]}, \bibinfo{author}{Mei\xfnm[ X.]},
  \bibinfo{author}{Huang\xfnm[ M.]}, \bibinfo{author}{Yang\xfnm[ Y.]},
  \bibinfo{author}{Fayad\xfnm[ Z.A.]}, \bibinfo{author}{Zhang\xfnm[ N.]},
  \bibinfo{author}{Diao\xfnm[ K.]}, \bibinfo{author}{Lin\xfnm[ B.]},
  \bibinfo{author}{Zhu\xfnm[ X.]}, \bibinfo{author}{Li\xfnm[ K.]}, et~al.
\newblock \bibinfo{title}{Chest ct findings in coronavirus disease-19
  (covid-19): relationship to duration of infection}.
\newblock \bibinfo{journal}{Radiology}
  \bibinfo{year}{2020};:\bibinfo{pages}{200463}.
\bibitem[{Butt et~al.(2020)Butt, Gill, Chun and Babu}]{butt2020deep}
\bibinfo{author}{Butt\xfnm[ C.]}, \bibinfo{author}{Gill\xfnm[ J.]},
  \bibinfo{author}{Chun\xfnm[ D.]}, \bibinfo{author}{Babu\xfnm[ B.A.]}.
\newblock \bibinfo{title}{Deep learning system to screen coronavirus disease
  2019 pneumonia}.
\newblock \bibinfo{journal}{Applied Intelligence}
  \bibinfo{year}{2020};:\bibinfo{pages}{1}.
\bibitem[{Chen et~al.(2020{\natexlab{a}})Chen, Wu, Zhang, Zhang, Gong, Zhao,
  Hu, Wang, Hu, Zheng et~al.}]{chen2020deep}
\bibinfo{author}{Chen\xfnm[ J.]}, \bibinfo{author}{Wu\xfnm[ L.]},
  \bibinfo{author}{Zhang\xfnm[ J.]}, \bibinfo{author}{Zhang\xfnm[ L.]},
  \bibinfo{author}{Gong\xfnm[ D.]}, \bibinfo{author}{Zhao\xfnm[ Y.]},
  \bibinfo{author}{Hu\xfnm[ S.]}, \bibinfo{author}{Wang\xfnm[ Y.]},
  \bibinfo{author}{Hu\xfnm[ X.]}, \bibinfo{author}{Zheng\xfnm[ B.]}, et~al.
\newblock \bibinfo{title}{Deep learning-based model for detecting 2019 novel
  coronavirus pneumonia on high-resolution computed tomography: a prospective
  study}.
\newblock \bibinfo{journal}{medRxiv} \bibinfo{year}{2020}{\natexlab{a}};.
\bibitem[{Chen et~al.(2020{\natexlab{b}})Chen, Kornblith, Norouzi and
  Hinton}]{chen2020simple}
\bibinfo{author}{Chen\xfnm[ T.]}, \bibinfo{author}{Kornblith\xfnm[ S.]},
  \bibinfo{author}{Norouzi\xfnm[ M.]}, \bibinfo{author}{Hinton\xfnm[ G.]}.
\newblock \bibinfo{title}{A simple framework for contrastive learning of visual
  representations}.
\newblock \bibinfo{journal}{arXiv preprint arXiv:200205709}
  \bibinfo{year}{2020}{\natexlab{b}};.
\bibitem[{Chen et~al.(2020{\natexlab{c}})Chen, Fan, Girshick and
  He}]{chen2020improved}
\bibinfo{author}{Chen\xfnm[ X.]}, \bibinfo{author}{Fan\xfnm[ H.]},
  \bibinfo{author}{Girshick\xfnm[ R.]}, \bibinfo{author}{He\xfnm[ K.]}.
\newblock \bibinfo{title}{Improved baselines with momentum contrastive
  learning}.
\newblock \bibinfo{journal}{arXiv preprint arXiv:200304297}
  \bibinfo{year}{2020}{\natexlab{c}};.
\bibitem[{Chen et~al.(2020{\natexlab{d}})Chen, Yao and
  Zhang}]{chen2020residual}
\bibinfo{author}{Chen\xfnm[ X.]}, \bibinfo{author}{Yao\xfnm[ L.]},
  \bibinfo{author}{Zhang\xfnm[ Y.]}.
\newblock \bibinfo{title}{Residual attention u-net for automated multi-class
  segmentation of covid-19 chest ct images}.
\newblock \bibinfo{journal}{arXiv preprint arXiv:200405645}
  \bibinfo{year}{2020}{\natexlab{d}};.
\bibitem[{Doersch and Zisserman(2017)}]{doersch2017multi}
\bibinfo{author}{Doersch\xfnm[ C.]}, \bibinfo{author}{Zisserman\xfnm[ A.]}.
\newblock \bibinfo{title}{Multi-task self-supervised visual learning}.
\newblock In: \bibinfo{booktitle}{Proceedings of the IEEE International
  Conference on Computer Vision}. \bibinfo{year}{2017}. p.
  \bibinfo{pages}{2051--2060}.
\bibitem[{Donahue et~al.(2014)Donahue, Jia, Vinyals, Hoffman, Zhang, Tzeng and
  Darrell}]{donahue2014decaf}
\bibinfo{author}{Donahue\xfnm[ J.]}, \bibinfo{author}{Jia\xfnm[ Y.]},
  \bibinfo{author}{Vinyals\xfnm[ O.]}, \bibinfo{author}{Hoffman\xfnm[ J.]},
  \bibinfo{author}{Zhang\xfnm[ N.]}, \bibinfo{author}{Tzeng\xfnm[ E.]},
  \bibinfo{author}{Darrell\xfnm[ T.]}.
\newblock \bibinfo{title}{Decaf: A deep convolutional activation feature for
  generic visual recognition}.
\newblock In: \bibinfo{booktitle}{International conference on machine
  learning}. \bibinfo{year}{2014}. p. \bibinfo{pages}{647--655}.
\bibitem[{Donahue and Simonyan(2019)}]{donahue2019large}
\bibinfo{author}{Donahue\xfnm[ J.]}, \bibinfo{author}{Simonyan\xfnm[ K.]}.
\newblock \bibinfo{title}{Large scale adversarial representation learning}.
\newblock In: \bibinfo{booktitle}{Advances in Neural Information Processing
  Systems}. \bibinfo{year}{2019}. p. \bibinfo{pages}{10541--10551}.
\bibitem[{Eaton-Rosen et~al.(2018)Eaton-Rosen, Bragman, Ourselin and
  Cardoso}]{eaton2018improving}
\bibinfo{author}{Eaton-Rosen\xfnm[ Z.]}, \bibinfo{author}{Bragman\xfnm[ F.]},
  \bibinfo{author}{Ourselin\xfnm[ S.]}, \bibinfo{author}{Cardoso\xfnm[ M.J.]}.
\newblock \bibinfo{title}{Improving data augmentation for medical image
  segmentation} \bibinfo{year}{2018};.
\bibitem[{Gozes et~al.(2020)Gozes, Frid-Adar, Greenspan, Browning, Zhang, Ji,
  Bernheim and Siegel}]{gozes2020rapid}
\bibinfo{author}{Gozes\xfnm[ O.]}, \bibinfo{author}{Frid-Adar\xfnm[ M.]},
  \bibinfo{author}{Greenspan\xfnm[ H.]}, \bibinfo{author}{Browning\xfnm[
  P.D.]}, \bibinfo{author}{Zhang\xfnm[ H.]}, \bibinfo{author}{Ji\xfnm[ W.]},
  \bibinfo{author}{Bernheim\xfnm[ A.]}, \bibinfo{author}{Siegel\xfnm[ E.]}.
\newblock \bibinfo{title}{Rapid ai development cycle for the coronavirus
  (covid-19) pandemic: Initial results for automated detection \& patient
  monitoring using deep learning ct image analysis}.
\newblock \bibinfo{journal}{arXiv preprint arXiv:200305037}
  \bibinfo{year}{2020};.
\bibitem[{Hadsell et~al.(2006)Hadsell, Chopra and
  LeCun}]{hadsell2006dimensionality}
\bibinfo{author}{Hadsell\xfnm[ R.]}, \bibinfo{author}{Chopra\xfnm[ S.]},
  \bibinfo{author}{LeCun\xfnm[ Y.]}.
\newblock \bibinfo{title}{Dimensionality reduction by learning an invariant
  mapping}.
\newblock In: \bibinfo{booktitle}{2006 IEEE Computer Society Conference on
  Computer Vision and Pattern Recognition (CVPR'06)}.
  \bibinfo{organization}{IEEE}; volume~\bibinfo{volume}{2};
  \bibinfo{year}{2006}. p. \bibinfo{pages}{1735--1742}.
\bibitem[{He et~al.(2019)He, Fan, Wu, Xie and Girshick}]{he2019momentum}
\bibinfo{author}{He\xfnm[ K.]}, \bibinfo{author}{Fan\xfnm[ H.]},
  \bibinfo{author}{Wu\xfnm[ Y.]}, \bibinfo{author}{Xie\xfnm[ S.]},
  \bibinfo{author}{Girshick\xfnm[ R.]}.
\newblock \bibinfo{title}{Momentum contrast for unsupervised visual
  representation learning}.
\newblock \bibinfo{journal}{arXiv preprint arXiv:191105722}
  \bibinfo{year}{2019};.
\bibitem[{He et~al.(2015)He, Zhang, Ren and Sun}]{he2015delving}
\bibinfo{author}{He\xfnm[ K.]}, \bibinfo{author}{Zhang\xfnm[ X.]},
  \bibinfo{author}{Ren\xfnm[ S.]}, \bibinfo{author}{Sun\xfnm[ J.]}.
\newblock \bibinfo{title}{Delving deep into rectifiers: Surpassing human-level
  performance on imagenet classification}.
\newblock In: \bibinfo{booktitle}{Proceedings of the IEEE international
  conference on computer vision}. \bibinfo{year}{2015}. p.
  \bibinfo{pages}{1026--1034}.
\bibitem[{He et~al.(2016)He, Zhang, Ren and Sun}]{he2016deep}
\bibinfo{author}{He\xfnm[ K.]}, \bibinfo{author}{Zhang\xfnm[ X.]},
  \bibinfo{author}{Ren\xfnm[ S.]}, \bibinfo{author}{Sun\xfnm[ J.]}.
\newblock \bibinfo{title}{Deep residual learning for image recognition}.
\newblock In: \bibinfo{booktitle}{Proceedings of the IEEE conference on
  computer vision and pattern recognition}. \bibinfo{year}{2016}. p.
  \bibinfo{pages}{770--778}.
\bibitem[{He et~al.(2020)He, Yang, Zhang, Zhao, Zhang, Xing and
  Xie}]{he2020sample}
\bibinfo{author}{He\xfnm[ X.]}, \bibinfo{author}{Yang\xfnm[ X.]},
  \bibinfo{author}{Zhang\xfnm[ S.]}, \bibinfo{author}{Zhao\xfnm[ J.]},
  \bibinfo{author}{Zhang\xfnm[ Y.]}, \bibinfo{author}{Xing\xfnm[ E.]},
  \bibinfo{author}{Xie\xfnm[ P.]}.
\newblock \bibinfo{title}{Sample-efficient deep learning for covid-19 diagnosis
  based on ct scans}.
\newblock \bibinfo{journal}{medRxiv} \bibinfo{year}{2020};.
\bibitem[{H{\'e}naff et~al.(2019)H{\'e}naff, Srinivas, De~Fauw, Razavi,
  Doersch, Eslami and Oord}]{henaff2019data}
\bibinfo{author}{H{\'e}naff\xfnm[ O.J.]}, \bibinfo{author}{Srinivas\xfnm[ A.]},
  \bibinfo{author}{De~Fauw\xfnm[ J.]}, \bibinfo{author}{Razavi\xfnm[ A.]},
  \bibinfo{author}{Doersch\xfnm[ C.]}, \bibinfo{author}{Eslami\xfnm[ S.]},
  \bibinfo{author}{Oord\xfnm[ A.v.d.]}.
\newblock \bibinfo{title}{Data-efficient image recognition with contrastive
  predictive coding}.
\newblock \bibinfo{journal}{arXiv preprint arXiv:190509272}
  \bibinfo{year}{2019};.
\bibitem[{Hjelm et~al.(2018)Hjelm, Fedorov, Lavoie-Marchildon, Grewal, Bachman,
  Trischler and Bengio}]{hjelm2018learning}
\bibinfo{author}{Hjelm\xfnm[ R.D.]}, \bibinfo{author}{Fedorov\xfnm[ A.]},
  \bibinfo{author}{Lavoie-Marchildon\xfnm[ S.]}, \bibinfo{author}{Grewal\xfnm[
  K.]}, \bibinfo{author}{Bachman\xfnm[ P.]}, \bibinfo{author}{Trischler\xfnm[
  A.]}, \bibinfo{author}{Bengio\xfnm[ Y.]}.
\newblock \bibinfo{title}{Learning deep representations by mutual information
  estimation and maximization}.
\newblock \bibinfo{journal}{arXiv preprint arXiv:180806670}
  \bibinfo{year}{2018};.
\bibitem[{Huang et~al.(2020)Huang, Wang, Li, Ren, Zhao, Hu, Zhang, Fan, Xu, Gu
  et~al.}]{huang2020clinical}
\bibinfo{author}{Huang\xfnm[ C.]}, \bibinfo{author}{Wang\xfnm[ Y.]},
  \bibinfo{author}{Li\xfnm[ X.]}, \bibinfo{author}{Ren\xfnm[ L.]},
  \bibinfo{author}{Zhao\xfnm[ J.]}, \bibinfo{author}{Hu\xfnm[ Y.]},
  \bibinfo{author}{Zhang\xfnm[ L.]}, \bibinfo{author}{Fan\xfnm[ G.]},
  \bibinfo{author}{Xu\xfnm[ J.]}, \bibinfo{author}{Gu\xfnm[ X.]}, et~al.
\newblock \bibinfo{title}{Clinical features of patients infected with 2019
  novel coronavirus in wuhan, china}.
\newblock \bibinfo{journal}{The Lancet}
  \bibinfo{year}{2020};\bibinfo{volume}{395}(\bibinfo{number}{10223}):\bibinfo{pages}{497--506}.
\bibitem[{Huang et~al.(2017)Huang, Liu, Van Der~Maaten and
  Weinberger}]{huang2017densely}
\bibinfo{author}{Huang\xfnm[ G.]}, \bibinfo{author}{Liu\xfnm[ Z.]},
  \bibinfo{author}{Van Der~Maaten\xfnm[ L.]}, \bibinfo{author}{Weinberger\xfnm[
  K.Q.]}.
\newblock \bibinfo{title}{Densely connected convolutional networks}.
\newblock In: \bibinfo{booktitle}{Proceedings of the IEEE conference on
  computer vision and pattern recognition}. \bibinfo{year}{2017}. p.
  \bibinfo{pages}{4700--4708}.
\bibitem[{Li et~al.(2020)Li, Qin, Xu, Yin, Wang, Kong, Bai, Lu, Fang, Song
  et~al.}]{li2020artificial}
\bibinfo{author}{Li\xfnm[ L.]}, \bibinfo{author}{Qin\xfnm[ L.]},
  \bibinfo{author}{Xu\xfnm[ Z.]}, \bibinfo{author}{Yin\xfnm[ Y.]},
  \bibinfo{author}{Wang\xfnm[ X.]}, \bibinfo{author}{Kong\xfnm[ B.]},
  \bibinfo{author}{Bai\xfnm[ J.]}, \bibinfo{author}{Lu\xfnm[ Y.]},
  \bibinfo{author}{Fang\xfnm[ Z.]}, \bibinfo{author}{Song\xfnm[ Q.]}, et~al.
\newblock \bibinfo{title}{Artificial intelligence distinguishes covid-19 from
  community acquired pneumonia on chest ct}.
\newblock \bibinfo{journal}{Radiology}
  \bibinfo{year}{2020};:\bibinfo{pages}{200905}.
\bibitem[{Li and Xia(2020)}]{li2020coronavirus}
\bibinfo{author}{Li\xfnm[ Y.]}, \bibinfo{author}{Xia\xfnm[ L.]}.
\newblock \bibinfo{title}{Coronavirus disease 2019 (covid-19): Role of chest ct
  in diagnosis and management}.
\newblock \bibinfo{journal}{American Journal of Roentgenology}
  \bibinfo{year}{2020};:\bibinfo{pages}{1--7}.
\bibitem[{Long et~al.(2020)Long, Xu, Shen, Zhang, Fan, Wang, Zeng, Li, Li and
  Li}]{long2020diagnosis}
\bibinfo{author}{Long\xfnm[ C.]}, \bibinfo{author}{Xu\xfnm[ H.]},
  \bibinfo{author}{Shen\xfnm[ Q.]}, \bibinfo{author}{Zhang\xfnm[ X.]},
  \bibinfo{author}{Fan\xfnm[ B.]}, \bibinfo{author}{Wang\xfnm[ C.]},
  \bibinfo{author}{Zeng\xfnm[ B.]}, \bibinfo{author}{Li\xfnm[ Z.]},
  \bibinfo{author}{Li\xfnm[ X.]}, \bibinfo{author}{Li\xfnm[ H.]}.
\newblock \bibinfo{title}{Diagnosis of the coronavirus disease (covid-19):
  rrt-pcr or ct?}
\newblock \bibinfo{journal}{European Journal of Radiology}
  \bibinfo{year}{2020};:\bibinfo{pages}{108961}.
\bibitem[{Mei et~al.(2020)Mei, Lee, Diao, Huang, Lin, Liu, Xie, Ma, Robson,
  Chung et~al.}]{mei2020artificial}
\bibinfo{author}{Mei\xfnm[ X.]}, \bibinfo{author}{Lee\xfnm[ H.C.]},
  \bibinfo{author}{Diao\xfnm[ K.y.]}, \bibinfo{author}{Huang\xfnm[ M.]},
  \bibinfo{author}{Lin\xfnm[ B.]}, \bibinfo{author}{Liu\xfnm[ C.]},
  \bibinfo{author}{Xie\xfnm[ Z.]}, \bibinfo{author}{Ma\xfnm[ Y.]},
  \bibinfo{author}{Robson\xfnm[ P.M.]}, \bibinfo{author}{Chung\xfnm[ M.]},
  et~al.
\newblock \bibinfo{title}{Artificial intelligence--enabled rapid diagnosis of
  patients with covid-19}.
\newblock \bibinfo{journal}{Nature Medicine}
  \bibinfo{year}{2020};:\bibinfo{pages}{1--5}.
\bibitem[{Narin et~al.(2020)Narin, Kaya and Pamuk}]{narin2020automatic}
\bibinfo{author}{Narin\xfnm[ A.]}, \bibinfo{author}{Kaya\xfnm[ C.]},
  \bibinfo{author}{Pamuk\xfnm[ Z.]}.
\newblock \bibinfo{title}{Automatic detection of coronavirus disease (covid-19)
  using x-ray images and deep convolutional neural networks}.
\newblock \bibinfo{journal}{arXiv:200310849} \bibinfo{year}{2020};.
\bibitem[{Oliveira et~al.(2017)Oliveira, Pereira and
  Silva}]{oliveira2017augmenting}
\bibinfo{author}{Oliveira\xfnm[ A.]}, \bibinfo{author}{Pereira\xfnm[ S.]},
  \bibinfo{author}{Silva\xfnm[ C.A.]}.
\newblock \bibinfo{title}{Augmenting data when training a cnn for retinal
  vessel segmentation: How to warp?}
\newblock In: \bibinfo{booktitle}{2017 IEEE 5th Portuguese Meeting on
  Bioengineering (ENBENG)}. \bibinfo{organization}{IEEE}; \bibinfo{year}{2017}.
  p. \bibinfo{pages}{1--4}.
\bibitem[{Oord et~al.(2018)Oord, Li and Vinyals}]{oord2018representation}
\bibinfo{author}{Oord\xfnm[ A.v.d.]}, \bibinfo{author}{Li\xfnm[ Y.]},
  \bibinfo{author}{Vinyals\xfnm[ O.]}.
\newblock \bibinfo{title}{Representation learning with contrastive predictive
  coding}.
\newblock \bibinfo{journal}{arXiv preprint arXiv:180703748}
  \bibinfo{year}{2018};.
\bibitem[{Parkhi et~al.(2012)Parkhi, Vedaldi, Zisserman and
  Jawahar}]{parkhi2012cats}
\bibinfo{author}{Parkhi\xfnm[ O.M.]}, \bibinfo{author}{Vedaldi\xfnm[ A.]},
  \bibinfo{author}{Zisserman\xfnm[ A.]}, \bibinfo{author}{Jawahar\xfnm[ C.]}.
\newblock \bibinfo{title}{Cats and dogs}.
\newblock In: \bibinfo{booktitle}{2012 IEEE conference on computer vision and
  pattern recognition}. \bibinfo{organization}{IEEE}; \bibinfo{year}{2012}. p.
  \bibinfo{pages}{3498--3505}.
\bibitem[{Paszke et~al.(2019)Paszke, Gross, Massa, Lerer, Bradbury, Chanan,
  Killeen, Lin, Gimelshein, Antiga et~al.}]{paszke2019pytorch}
\bibinfo{author}{Paszke\xfnm[ A.]}, \bibinfo{author}{Gross\xfnm[ S.]},
  \bibinfo{author}{Massa\xfnm[ F.]}, \bibinfo{author}{Lerer\xfnm[ A.]},
  \bibinfo{author}{Bradbury\xfnm[ J.]}, \bibinfo{author}{Chanan\xfnm[ G.]},
  \bibinfo{author}{Killeen\xfnm[ T.]}, \bibinfo{author}{Lin\xfnm[ Z.]},
  \bibinfo{author}{Gimelshein\xfnm[ N.]}, \bibinfo{author}{Antiga\xfnm[ L.]},
  et~al.
\newblock \bibinfo{title}{Pytorch: An imperative style, high-performance deep
  learning library}.
\newblock In: \bibinfo{booktitle}{Advances in Neural Information Processing
  Systems}. \bibinfo{year}{2019}. p. \bibinfo{pages}{8024--8035}.
\bibitem[{Pereira et~al.(2016)Pereira, Pinto, Alves and
  Silva}]{pereira2016brain}
\bibinfo{author}{Pereira\xfnm[ S.]}, \bibinfo{author}{Pinto\xfnm[ A.]},
  \bibinfo{author}{Alves\xfnm[ V.]}, \bibinfo{author}{Silva\xfnm[ C.A.]}.
\newblock \bibinfo{title}{Brain tumor segmentation using convolutional neural
  networks in mri images}.
\newblock \bibinfo{journal}{IEEE transactions on medical imaging}
  \bibinfo{year}{2016};\bibinfo{volume}{35}(\bibinfo{number}{5}):\bibinfo{pages}{1240--1251}.
\bibitem[{Salehi et~al.(2020)Salehi, Abedi, Balakrishnan and
  Gholamrezanezhad}]{salehi2020coronavirus}
\bibinfo{author}{Salehi\xfnm[ S.]}, \bibinfo{author}{Abedi\xfnm[ A.]},
  \bibinfo{author}{Balakrishnan\xfnm[ S.]},
  \bibinfo{author}{Gholamrezanezhad\xfnm[ A.]}.
\newblock \bibinfo{title}{Coronavirus disease 2019 {(COVID-19)}: a systematic
  review of imaging findings in 919 patients}.
\newblock \bibinfo{journal}{American Journal of Roentgenology}
  \bibinfo{year}{2020};:\bibinfo{pages}{1--7}.
\bibitem[{Selvaraju et~al.(2017)Selvaraju, Cogswell, Das, Vedantam, Parikh and
  Batra}]{selvaraju2017grad}
\bibinfo{author}{Selvaraju\xfnm[ R.R.]}, \bibinfo{author}{Cogswell\xfnm[ M.]},
  \bibinfo{author}{Das\xfnm[ A.]}, \bibinfo{author}{Vedantam\xfnm[ R.]},
  \bibinfo{author}{Parikh\xfnm[ D.]}, \bibinfo{author}{Batra\xfnm[ D.]}.
\newblock \bibinfo{title}{Grad-cam: Visual explanations from deep networks via
  gradient-based localization}.
\newblock In: \bibinfo{booktitle}{Proceedings of the IEEE international
  conference on computer vision}. \bibinfo{year}{2017}. p.
  \bibinfo{pages}{618--626}.
\bibitem[{Shan et~al.(2020)Shan, Gao, Wang, Shi, Shi, Han, Xue, Shen and
  Shi}]{shan+2020lung}
\bibinfo{author}{Shan\xfnm[ F.]}, \bibinfo{author}{Gao\xfnm[ Y.]},
  \bibinfo{author}{Wang\xfnm[ J.]}, \bibinfo{author}{Shi\xfnm[ W.]},
  \bibinfo{author}{Shi\xfnm[ N.]}, \bibinfo{author}{Han\xfnm[ M.]},
  \bibinfo{author}{Xue\xfnm[ Z.]}, \bibinfo{author}{Shen\xfnm[ D.]},
  \bibinfo{author}{Shi\xfnm[ Y.]}.
\newblock \bibinfo{title}{Lung infection quantification of covid-19 in ct
  images with deep learning}.
\newblock \bibinfo{journal}{arXiv:200304655} \bibinfo{year}{2020};.
\bibitem[{Shi et~al.(2020{\natexlab{a}})Shi, Wang, Shi, Wu, Wang, Tang, He, Shi
  and Shen}]{shi2020review}
\bibinfo{author}{Shi\xfnm[ F.]}, \bibinfo{author}{Wang\xfnm[ J.]},
  \bibinfo{author}{Shi\xfnm[ J.]}, \bibinfo{author}{Wu\xfnm[ Z.]},
  \bibinfo{author}{Wang\xfnm[ Q.]}, \bibinfo{author}{Tang\xfnm[ Z.]},
  \bibinfo{author}{He\xfnm[ K.]}, \bibinfo{author}{Shi\xfnm[ Y.]},
  \bibinfo{author}{Shen\xfnm[ D.]}.
\newblock \bibinfo{title}{Review of artificial intelligence techniques in
  imaging data acquisition, segmentation and diagnosis for covid-19}.
\newblock \bibinfo{journal}{arXiv:200402731}
  \bibinfo{year}{2020}{\natexlab{a}};.
\bibitem[{Shi et~al.(2020{\natexlab{b}})Shi, Xia, Shan, Wu, Wei, Yuan, Jiang,
  Gao, Sui and Shen}]{shi2020large}
\bibinfo{author}{Shi\xfnm[ F.]}, \bibinfo{author}{Xia\xfnm[ L.]},
  \bibinfo{author}{Shan\xfnm[ F.]}, \bibinfo{author}{Wu\xfnm[ D.]},
  \bibinfo{author}{Wei\xfnm[ Y.]}, \bibinfo{author}{Yuan\xfnm[ H.]},
  \bibinfo{author}{Jiang\xfnm[ H.]}, \bibinfo{author}{Gao\xfnm[ Y.]},
  \bibinfo{author}{Sui\xfnm[ H.]}, \bibinfo{author}{Shen\xfnm[ D.]}.
\newblock \bibinfo{title}{Large-scale screening of covid-19 from community
  acquired pneumonia using infection size-aware classification}.
\newblock \bibinfo{journal}{arXiv preprint arXiv:200309860}
  \bibinfo{year}{2020}{\natexlab{b}};.
\bibitem[{Snell et~al.(2017)Snell, Swersky and Zemel}]{snell2017prototypical}
\bibinfo{author}{Snell\xfnm[ J.]}, \bibinfo{author}{Swersky\xfnm[ K.]},
  \bibinfo{author}{Zemel\xfnm[ R.]}.
\newblock \bibinfo{title}{Prototypical networks for few-shot learning}.
\newblock In: \bibinfo{booktitle}{Advances in neural information processing
  systems}. \bibinfo{year}{2017}. p. \bibinfo{pages}{4077--4087}.
\bibitem[{Sohn(2016)}]{sohn2016improved}
\bibinfo{author}{Sohn\xfnm[ K.]}.
\newblock \bibinfo{title}{Improved deep metric learning with multi-class n-pair
  loss objective}.
\newblock In: \bibinfo{booktitle}{Advances in neural information processing
  systems}. \bibinfo{year}{2016}. p. \bibinfo{pages}{1857--1865}.
\bibitem[{Vinyals et~al.(2016)Vinyals, Blundell, Lillicrap, Wierstra
  et~al.}]{vinyals2016matching}
\bibinfo{author}{Vinyals\xfnm[ O.]}, \bibinfo{author}{Blundell\xfnm[ C.]},
  \bibinfo{author}{Lillicrap\xfnm[ T.]}, \bibinfo{author}{Wierstra\xfnm[ D.]},
  et~al.
\newblock \bibinfo{title}{Matching networks for one shot learning}.
\newblock In: \bibinfo{booktitle}{Advances in neural information processing
  systems}. \bibinfo{year}{2016}. p. \bibinfo{pages}{3630--3638}.
\bibitem[{Wang et~al.(2020)Wang, Wang, Ye and Liu}]{wang2020review}
\bibinfo{author}{Wang\xfnm[ L.s.]}, \bibinfo{author}{Wang\xfnm[ Y.r.]},
  \bibinfo{author}{Ye\xfnm[ D.w.]}, \bibinfo{author}{Liu\xfnm[ Q.q.]}.
\newblock \bibinfo{title}{A review of the 2019 novel coronavirus (covid-19)
  based on current evidence}.
\newblock \bibinfo{journal}{International Journal of Antimicrobial Agents}
  \bibinfo{year}{2020};:\bibinfo{pages}{105948}.
\bibitem[{Wu et~al.(2018)Wu, Xiong, Yu and Lin}]{wu2018unsupervised}
\bibinfo{author}{Wu\xfnm[ Z.]}, \bibinfo{author}{Xiong\xfnm[ Y.]},
  \bibinfo{author}{Yu\xfnm[ S.X.]}, \bibinfo{author}{Lin\xfnm[ D.]}.
\newblock \bibinfo{title}{Unsupervised feature learning via non-parametric
  instance discrimination}.
\newblock In: \bibinfo{booktitle}{Proceedings of the IEEE Conference on
  Computer Vision and Pattern Recognition}. \bibinfo{year}{2018}. p.
  \bibinfo{pages}{3733--3742}.
\bibitem[{Yan et~al.(2018)Yan, Wang, Lu, Zhang, Harrison, Bagheri and
  Summers}]{yan2018deep}
\bibinfo{author}{Yan\xfnm[ K.]}, \bibinfo{author}{Wang\xfnm[ X.]},
  \bibinfo{author}{Lu\xfnm[ L.]}, \bibinfo{author}{Zhang\xfnm[ L.]},
  \bibinfo{author}{Harrison\xfnm[ A.P.]}, \bibinfo{author}{Bagheri\xfnm[ M.]},
  \bibinfo{author}{Summers\xfnm[ R.M.]}.
\newblock \bibinfo{title}{Deep lesion graphs in the wild: relationship learning
  and organization of significant radiology image findings in a diverse
  large-scale lesion database}.
\newblock In: \bibinfo{booktitle}{Proceedings of the IEEE Conference on
  Computer Vision and Pattern Recognition}. \bibinfo{year}{2018}. p.
  \bibinfo{pages}{9261--9270}.
\bibitem[{Zhao et~al.(2019)Zhao, Balakrishnan, Durand, Guttag and
  Dalca}]{zhao2019data}
\bibinfo{author}{Zhao\xfnm[ A.]}, \bibinfo{author}{Balakrishnan\xfnm[ G.]},
  \bibinfo{author}{Durand\xfnm[ F.]}, \bibinfo{author}{Guttag\xfnm[ J.V.]},
  \bibinfo{author}{Dalca\xfnm[ A.V.]}.
\newblock \bibinfo{title}{Data augmentation using learned transformations for
  one-shot medical image segmentation}.
\newblock In: \bibinfo{booktitle}{Proceedings of the IEEE conference on
  computer vision and pattern recognition}. \bibinfo{year}{2019}. p.
  \bibinfo{pages}{8543--8553}.
\bibitem[{Zhao et~al.(2020)Zhao, Zhang, He and Xie}]{zhao2020COVID-CT-Dataset}
\bibinfo{author}{Zhao\xfnm[ J.]}, \bibinfo{author}{Zhang\xfnm[ Y.]},
  \bibinfo{author}{He\xfnm[ X.]}, \bibinfo{author}{Xie\xfnm[ P.]}.
\newblock \bibinfo{title}{Covid-ct-dataset: a ct scan dataset about covid-19}.
\newblock \bibinfo{journal}{arXiv preprint arXiv:200313865}
  \bibinfo{year}{2020};.
\bibitem[{Zheng et~al.(2020)Zheng, Deng, Fu, Zhou, Feng, Ma, Liu and
  Wang}]{zheng2020deep}
\bibinfo{author}{Zheng\xfnm[ C.]}, \bibinfo{author}{Deng\xfnm[ X.]},
  \bibinfo{author}{Fu\xfnm[ Q.]}, \bibinfo{author}{Zhou\xfnm[ Q.]},
  \bibinfo{author}{Feng\xfnm[ J.]}, \bibinfo{author}{Ma\xfnm[ H.]},
  \bibinfo{author}{Liu\xfnm[ W.]}, \bibinfo{author}{Wang\xfnm[ X.]}.
\newblock \bibinfo{title}{Deep learning-based detection for covid-19 from chest
  ct using weak label}.
\newblock \bibinfo{journal}{medRxiv} \bibinfo{year}{2020};.

\end{thebibliography}

\end{document}